\documentstyle[preprint,aps,psfig,epsf,rotate]{revtex}

\begin{document}
\newcommand{\sfrac}[2]{\mbox{\footnotesize $\frac{#1}{#2}$}}
\newcommand{\dcsb}{D$\chi$SB} \def\psibar{{\overline{\psi}}}

\tightenlines

\preprint{ \vbox{ \hbox{ADP-01-19/T454} }}

\title{Dependence of the Chiral Symmetry Restoration Transition on the
  Quark Self-Energy Kernel} \author{A.~Bender, W.~Detmold and
  A.W.~Thomas} \address{Special Research Centre for the Subatomic
  Structure of Matter, and Department of Physics and Mathematical
  Physics, Adelaide University, 5005, Australia}
  
\maketitle

\begin{abstract}
  The dependence of the dressed quark propagator on the quark chemical
  potential is investigated in various models based on the
  Dyson-Schwinger equations. We find that the critical chemical
  potential of the chiral symmetry restoration transitions is strongly
  dependent on the nature of the interaction kernel in the infrared
  region.
\end{abstract}
\vspace{1cm}


The study of the thermodynamic phase structure of QCD has many
implications for astrophysics and relativistic heavy ion collisions.
An understanding of the behaviour of strongly interacting matter in
regions of non-zero temperature and density will give insight into the
nature of primordial baryogenesis and the dynamics of neutron stars.
Heavy ion colliders such as RHIC and LHC will soon provide
experimental data in thermodynamic regions where chiral symmetry is
predicted to be restored.

The structure of QCD at non-zero temperature and zero density has been
explored extensively using numerical simulations of lattice QCD. These
results and other model based calculations suggest that there is a
critical temperature, $T_{\rm crit}\simeq170$--190~MeV
\cite{Karsch:2000vy}, above which normal nuclear matter undergoes a
phase transition to a quark-gluon plasma in which chiral symmetry is
restored. Unfortunately, current lattice simulation methods have
significant difficulties when a non-zero quark or baryon density is
introduced \cite{Barbour:1998ej}. For this reason, studies in models
that implement the pertinent features of QCD are important.  Here, we
present results from such a study utilising the Dyson-Schwinger
equation (DSE) framework \cite{DSErefs}.

At present we make no attempt to incorporate the colour
superconducting ground state that has been shown to occur at
asymptotic quark densities \cite{cscrefs}. The
relevance of these results to QCD at densities of interest in nuclear
physics (where perturbative results are suspect) is not well
understood.  These densities have only been investigated in simple
approximations such as the Nambu--Jona-Lasinio model \cite{alford98}
and the Munczek-Nemirovsky (DSE) model \cite{Bloch:1999vk} which may
not be reliable representations of QCD in this regime.  An extension
of the current work may address these deficiencies and is under
investigation.

In principle, the DSE's determine all possible information about a
quantum field theory by providing all of its Schwinger functions.
Unfortunately, they form a countably infinite set of coupled integral
equations with the equation for an $n$--point Schwinger function
depending on $(n+1)$ and higher point functions. Our approach is
derived from these equations by specifying a scheme for truncating
this infinite system through modelling the requisite, undetermined
Schwinger functions.
 
Previously, simple DSE models \cite{Bender:1998jf,Bloch:1999vk} have
been used to investigate the restoration of chiral symmetry as the
quark density (or chemical potential) increases. A phase transition
has been found at a critical chemical potential, $\mu_{\rm
  crit}\sim$~300--400 MeV. Here we examine the effect on this result
of using different quark self-energy kernels.


In Euclidean space (with metric $g^{\mu\nu}=\delta^{\mu\nu}$, and
Clifford algebra $\{\gamma_\mu,\gamma_\nu\}=2\delta_{\mu\nu}$,
$\gamma_5=-\gamma_1\gamma_2\gamma_3\gamma_4$), we separate the inverse
quark propagator, $S$, at non-zero quark chemical potential, $\mu$,
into its spacelike and timelike vector parts, $A$ and $C$, and scalar
part, $B$:
\begin{eqnarray} 
S^{-1}(\tilde p;\zeta) & \doteq & i
  \vec{\gamma}\cdot \vec{\bf p} A(\tilde p ;\zeta ) +
  i \gamma_4 \omega_p C(\tilde p;\zeta ) + B(\tilde p;\zeta)
\label{invprop}
\end{eqnarray}
where the quark momentum $\tilde p\equiv(\vec{\bf
  p},\omega_p=p_4+i\mu)$.  The propagator is renormalised at the (O(4)
invariant) scale $\zeta$ which is large enough that
\begin{equation}
\label{freeprop}
\left.S^{-1}(\tilde p;\zeta)\right|_{\tilde p^2=\zeta^2}^{\mu=0}=
S^{-1}_0(\tilde p;\zeta)=
i\gamma\cdot p +m_R(\zeta)\,,
\end{equation}
where $S_0$ is the free propagator for a quark with running current
mass $m_R(\zeta)$.  In terms of the regularised self-energy
$\Sigma^\prime(\tilde p,\Lambda)$, the inverse propagator can be
expressed as
\begin{eqnarray} 
S^{-1}(\tilde p;\zeta)   & = &
Z_2^A i \vec{\gamma} \cdot \vec{\bf p} +
Z_2 (i\gamma_4 \omega_p + m_{\rm bare}(\Lambda)) +
\Sigma^\prime(\tilde p ,\Lambda )\,,  
\label{quarkdse}
\end{eqnarray} 
where $m_{\rm bare}(\Lambda)$ is the bare, regularisation dependent
mass occuring in the Lagrangian, $\Lambda$ a regularisation parameter
(we use $\int_k^\Lambda=\int_0^\Lambda \frac{k^3\,dk}{(2\pi)^4}\int
d\Omega_3$ with a simple cutoff regulator $\Lambda$), and $Z_2^A$ and
$Z_2$ are respectively, the spacelike and timelike field
renormalisations, dependent on both the regularisation and
renormalisation scales. As we renormalise at $\mu=0$ where O(4)
invariance is preserved, $Z_2^A=Z_2$.

The DSE satisfied by the self-energy is
\begin{equation}
\Sigma^\prime(\tilde p,\Lambda)=g^2\int_k^\Lambda D_{\mu\nu}^{ab}(p-k)
t^a\gamma_\mu S(\tilde k)\Gamma_\nu^b(p,k)\,,
\end{equation}
where $D_{\mu\nu}^{ab}(q)$ and $\Gamma_\nu^b(p,k)$ are the full,
nonperturbative gluon propagator and the quark gluon vertex
respectively and \{a,b\} are colour indices with
$t^a=\frac{\lambda^a}{2}$ for the standard Gell-Mann SU(3)
representation $\lambda^a$ \cite{DSErefs}.

As is the case for the propagator, the self-energy can be expressed in
terms of three scalar functions $\Sigma^\prime_A$, $\Sigma^\prime_B$
and $\Sigma^\prime_C$ as
\begin{equation}
\Sigma^\prime(\tilde p;\Lambda)=i\vec{\gamma} \cdot 
\vec{\bf p}\Sigma^\prime_A(\tilde p;\Lambda) 
+ i\gamma_4 \omega_p\Sigma^\prime_C(\tilde p;\Lambda)+
\Sigma^\prime_B(\tilde p;\Lambda).
\end{equation}
These functions satisfy the coupled DSEs,
\begin{eqnarray}
\label{selfdse}
\Sigma_X^\prime(\tilde p;\Lambda)
&=&  \int^\Lambda_k g^2 D_{\mu\nu}^{ab}(\tilde p-\tilde k)\,
\sfrac{1}{4}{\rm tr}\left[{\cal P}_X \gamma_\mu t^aS(\tilde k)
\Gamma^b_\nu(\tilde p,\tilde k)\right]\,,
\end{eqnarray}
where $X= A$, $B$, $C$; ${\cal P}_A = - (Z_1^Ai \vec{\gamma}\cdot
\vec{\bf p}/|\vec{\bf p}|^2) $, ${\cal P}_B = Z_1$, ${\cal P}_C=
-(Z_1i \gamma_4/\omega_p)$ and the trace is over Dirac and colour
space.  We constrain the vertex renormalisation $Z_1$ to be the same
as the field renormalisation, $Z_1=Z_2$, which is consistent with the
use of the bare vertex $\Gamma^b_\nu(\tilde q,\tilde p)=\gamma_\nu
t^b$ \cite{DSErefs}.

The renormalisation condition (\ref{freeprop}) implies that
\begin{eqnarray}
Z_2(\zeta^2,\Lambda^2)&=&1-\left.\Sigma^\prime_A(\zeta,\Lambda)
\right|_{\mu=0} \\
m_R(\zeta)&=&Z_2(\zeta^2,\Lambda^2)m_{\rm bare}(\Lambda) +
\left.\Sigma^\prime_B(\zeta,\Lambda)\right|_{\mu=0}. \nonumber 
\end{eqnarray}
Thus the regulator independent propagator functions are defined by
\begin{eqnarray}
\label{propfunc}
A(\tilde p;\zeta)=1+\Sigma^\prime_A(\tilde p;\Lambda)-\left.
  \Sigma^\prime_A(\tilde p;\Lambda)\right|^{\mu=0}_{\tilde
  p^2=\zeta^2}
\nonumber \\
B(\tilde p;\zeta)=m_R(\zeta) +\Sigma^\prime_B(\tilde p;\Lambda)-\left.
  \Sigma^\prime_B(\tilde p;\Lambda)\right|^{\mu=0}_{\tilde
  p^2=\zeta^2}
\\
C(\tilde p;\zeta)=1+\Sigma^\prime_C(\tilde
p;\Lambda)-\left.\Sigma^\prime_C(\tilde
  p;\Lambda)\right|^{\mu=0}_{\tilde p^2=\zeta^2}. \nonumber 
\end{eqnarray}
Equations (\ref{invprop}) through (\ref{propfunc}) form a coupled set
of nonlinear integral equations for the quark propagator, and once the
self-energy kernel $D\Gamma$ is specified we can proceed to solve
them.


Here, we consider various models for this kernel, and in particular,
separately for $D_{\mu\nu}^{ab}$ and $\Gamma^b_\mu$.  There are some
constraints on the models that we can use to guide our choice; the
main requirement is that they must be capable of producing dynamical
chiral symmetry breaking (\dcsb) and confinement \cite{Hawes:1998cw}
as this is what is found experimentally. The behaviour of these
objects in the ultraviolet region is also dictated by the results of
perturbation theory.

The full Euclidean space gluon propagator can be written in Landau
gauge as
\begin{equation}
  g^2D^{ab}_{\mu\nu}(q)=g^2\delta^{ab}D_{\mu\nu}(q)
  =\delta^{ab}\left(\delta_{\mu\nu}-\frac{q_\mu
        q_\nu}{q^2}\right) \frac{{\cal G}(q^2)}{q^2},
\end{equation} 
where we have assumed that the effect of quark chemical potential on
the gluon propagator through quark loops is small in comparison to
that on the quark propagator. To provide a model for it, we specify
$\frac{{\cal G}(q^2)}{q^2}$.  As the first example, we choose:
\begin{equation}
\label{MNprop}
\frac{{\cal G}(q^2)}{q^2}=(2\pi)^4{\cal G}\delta^4(q), 
\end{equation} 
which corresponds to the Munczek-Nemirovsky \cite{Munczek:1983dx} (MN)
model.  This model has been studied in the context of finite
temperature and density \cite{Bloch:1999vk,Blaschke:1998bj} and has
the advantage that the integral equations of the DSE reduce to
algebraic equations (albeit nonlinear).

The remaining models are variations on the form:
\begin{eqnarray}
\label{MRprop}
\frac{{\cal G}(q^2)}{q^2}=8\pi^4\Delta{\cal G}\delta^4(q)
+4\pi^2(2-\Delta)\frac{{\cal G}\,q^{2}}{\omega^6}e^{-\frac{q^2}{\omega^2}}
+\frac{4\pi^2\gamma_m\left[1-\exp\left(\frac{-q^2}{4m_t^2}\right)\right]}
{q^2(\sfrac{1}{2})\ln\left[\tau+\left(1+\frac{q^2}{\Lambda_{QCD}^2}\right)^2\right]}\,,
\end{eqnarray} 
where the one loop anomalous dimension
$\gamma_m=\frac{12}{11N_c-2N_f}$, the QCD scale parameter is set at
$\Lambda_{QCD}=0.275\,{\rm GeV}$, and $\tau=e^2-1$.  The last term in
Eq.(\ref{MRprop}) implements the results of perturbative QCD to one
loop.  However, the quantities we study in this paper are determined
primarily by the low momentum structure of the kernel and are
relatively insensitive to this ultra-violet behaviour.  The first two
terms of the model provide some infrared enhancement to the
self-energy, leading to dynamical chiral symmetry breaking (\dcsb).
The exact form of the IR enhancement of the quark self-energy kernel
is unknown \cite{Hawes:1998cw}\footnote{Indeed, recent lattice
  calculations \cite{Bonnet:2000kw} and DSE studies \cite{GluonDSE}
  suggest the enhancement may in fact not occur in the gluon
  propagator but through the vertex. In our approach, it is full
  kernel $D\Gamma$ that is modelled and the particular separation into
  gluon propagator and vertex function is arbitrary.}, and by varying
$\Delta$ and to some extent $\omega$ it is possible to explore various
possibilities.
For $\Delta=2$, the propagator (\ref{MRprop}) essentially reduces to
that considered in Refs.\cite{Bender:1998jf,Frank:1996uk}, however we
have also included the UV tail which provides the correct leading-log
asymptotic behaviour for the quark propagator.  For $\Delta=1$ this is
the propagator considered at zero density in Ref. \cite{Maris:1997tm}.
Finally, with $\Delta=0$ we get the propagator used in
Ref.\cite{Maris:1999nt}.

In order to motivate the different models used here for the vertex, we
first briefly outline the method used to obtain the critical chemical
potential. For a given self-energy kernel, the DSE's for the quark
propagator are solved self-consistently to find solutions representing
the Nambu-Goldstone phase (characterised by \dcsb) and a Wigner-Weyl
phase (corresponding to the quark-gluon plasma where chiral symmetry
is restored). The energetic stability of these two solutions is then
compared using an effective action for the composite operators $S$ and
$D$.

One commonly used truncation of the DSE (\ref{selfdse}), called the
rainbow approximation, involves replacing the full vertex in Eq.
(\ref{selfdse}) by the bare vertex,
\begin{equation}
\Gamma_\nu^a(k,p)=t_a\gamma_\nu.
\end{equation}
In this case, the correct action to use is the
Cornwall-Jackiw-Tomboulis (CJT) effective action
\cite{Cornwall:1974vz} which is given by
\begin{equation}
{\cal A}_{\rm CJT}[S] = {\rm TrLn}\left[S_0^{-1}S\right]
+\sfrac{1}{2}{\rm Tr}\left[\Sigma S\right].
\end{equation}

We wish to explore the effects of moving beyond the rainbow
approximation (ideally to a vertex that respects the various
symmetries of the full vertex, see Refs. \cite{DSErefs,BCCP}), but
must first address the question of which effective action should be
used to measure the energy of the different solutions.  For a more
complicated vertex, functionally dependent on the quark propagator,
the CJT action is not the correct object to use \cite{FUTURE}.
However the correction required is so far only calculable in a limited
number of cases\footnote{This is essentially the same problem that
  frustrates a consistent truncation of the DSE and corresponding BSE
  equations}. This restricts the types of models that we can use for
the vertex.

Consequently, we use the next level of truncation of the DSE
(\ref{selfdse}); a one loop vertex involving the non-perturbative
quark and gluon propagators. This vertex was introduced in
Ref.\cite{Bender:1996bb} and used further in Refs.
\cite{Bloch:1999vk,Hellstern:1997nv}.  In combination with the
appropriate Bethe-Salpeter kernel, it preserves the axial-vector Ward
identity and is systematically improvable.  The form for this vertex
is
\begin{eqnarray}
\label{oneloop}
\Gamma^{1L}_{\mu;a}(k,p)=t^a\Big[\gamma_\mu
  +\frac{1}{6}\int^\Lambda_lg^2 D_{\rho\sigma}((p-l)^2)\gamma_\rho
  S(l+k-p)\gamma_\mu S(l)\gamma_\sigma\Big] \,.
\end{eqnarray}
The correct action to use in combination with this vertex is,
\begin{equation}
{\cal A}_{\rm 1L}[S,D] = {\cal A}_{\rm CJT}[S] + {\cal A}_{\Gamma}[S,D],
\end{equation}
where \cite{FUTURE},
\begin{eqnarray}
  {\cal A}_\Gamma[S,D]=-\sfrac{1}{24}\int d^4pd^4qd^4r 
  D_{\rho\sigma}(p-q)D_{\mu\nu}(r-q) 
{\rm
  Tr}\left[\gamma^\mu S_p\gamma^\rho S_q\gamma^\nu S_r\gamma^\sigma
  S_{p-q+r}\right]. 
\end{eqnarray}


Having described the basis of the calculations, we now turn to the
most important results.  For the MN model with a bare vertex the
analytic solutions are known. In the chiral limit, the two relevant
solutions are:
\begin{eqnarray}
S_{1}:&\hspace{0.5cm}B(\tilde p)=0 \hspace{0.5cm} &A(\tilde
p)=\frac{1}{2}\left[1+\sqrt{1+\frac{8{\cal G}}{\tilde p^2}}\right]
\nonumber
\\
S_{2}:&\hspace{0.5cm}B(\tilde p)=2\sqrt{{\cal G}-\tilde
  p^2}\hspace{0.5cm} &A(\tilde p)=2\,.  
\end{eqnarray} 
In order to define the symmetric and \dcsb~solutions for the MN model
at non-zero quark chemical potential, it is necessary to provide a
prescription for the transition from solution $S_2$ to solution $S_1$
as $|\tilde p|\to\infty$. To generalise the $\mu=0$ case (where the
reality of B determines the transition point), we set $S_{\rm D\chi
  SB}= S_2$ for $\Re(\tilde p^2)<G+\mu^2$ and $S_{\rm D\chi SB}= S_1$
otherwise. We stress that while this prescription is a reasonable
generalisation of the model to finite $\mu$, it is essentially
arbitrary and different results are obtained with other prescriptions
\cite{FUTURE}.

For this model, the critical chemical potential ($\mu_{\rm crit}$ such
that $\Delta_{\rm CJT}\equiv{\cal A}_{\rm CJT}[S_{\rm D\chi SB}]-{\cal
  A}_{\rm CJT}[S_{\rm Symm.}]=0$) is displayed as a function of the
coupling strength ${\cal G}$ in Fig. 1. As $\cal G$ is increased,
there is a corresponding monotonic increase in $\mu_{\rm crit}$ which
can be fit exceedingly well with the simple form, $\mu_{\rm
  crit}({\cal G})=0.335\sqrt{\cal G}$. The ``physical'' value of the
model parameter $\cal G$ can be set by requiring that it yield the
correct zero density pion decay constant, $f_\pi$.  For this coupling,
the critical chemical potential is $\mu_{\rm crit}=340$~MeV.

For the more phenomenologically appropriate models using the
Maris-Roberts type effective quark-antiquark interactions, Eq.
(\ref{MRprop}), the integral equations (\ref{selfdse}) are solved
using an iterative procedure for $\zeta=19$~GeV and
$\Lambda\sim10^3$~GeV.  The quark condensate is extracted from the
asymptotic behaviour of the quark mass function
$M(p^2)=B(p^2;\zeta)/A(p^2;\zeta)$,
\begin{equation}
M(p^2)
\stackrel{p^2\to\infty}{\longrightarrow}\frac{2\pi\gamma_m}{N_c}
\frac{-\langle\psibar\psi\rangle^0}{p^2\left[\sfrac{1}{2}
\ln\left(\frac{p^2}{\Lambda_{QCD}^2}\right)\right]^{1-\gamma_m}}, 
\end{equation} 
which is accurate up to ${\cal O}(\frac{\Lambda_{QCD}^2}{\zeta^2})$.
The zero density pion mass and decay constant are calculated using
analytic approximations to the solutions of the corresponding
Bethe-Salpeter equation \cite{Frank:1996uk}. These values are used to
fix the bare quark mass and the parameters ${\cal G}$ and $\omega$ in
the gluon propagator.

Calculations of the critical chemical potential in these models give
qualitatively similar results to those of the MN model. Data from the
$\Delta=1$ and $\Delta=2$ cases is shown for a range of ${\cal G}$ in
the upper panel of Figure 1. It is interesting to explore the
parameter dependence of these models. Figures 2 and 3 present the
normalised CJT action differences $-\Delta_{\rm CJT}[\mu]/\Delta_{\rm
  CJT}[0]$ as a function of $\mu$ for variation of the three
parameters ${\cal G}$, $\Delta$ and $\omega$. The results for
different parameter choices are normalised to $-1$ at zero chemical
potential as $|\Delta_{\rm CJT}[0]|$ increases rapidly with $\cal G$.
For $\Delta=2$, Fig. 2 shows that as the total infrared strength $\cal
G$ is increased $\mu_{\rm crit}$ increases in a manner similar to that
of the MN model. Fig. 3 shows the normalised action differences for
variation with $\Delta$ (left panel, ${\cal G}=0.26$~GeV$^2$,
$\omega=0.3$~GeV) and $\omega$ (right panel, ${\cal G}=0.26$~GeV$^2$,
$\Delta=1$).  As $\Delta$ decreases from 2 (where the entire infrared
strength is in the delta function at zero momentum) to 0 (where the
effective quark-antiquark interaction vanishes at zero momentum), the
critical chemical potential decreases for a given coupling strength
and fixed value of $\omega$. Similarly, as the infrared strength of
the propagator is broadened by increasing $\omega$ for a given $\cal
G$ and $\Delta$, $\mu_{\rm crit}$ again decreases.  Table 1 summarises
the parameters used and the corresponding values of the critical
chemical potential for the various models considered here.

The effect on $\mu_{\rm crit}$ of using the one loop vertex is
twofold. First, the solutions themselves are modified, resulting in a
change to both the physical values for the model parameters and the
action differences. Using the corrected action, ${\cal A}_{\rm 1L}$,
and the MN gluon propagator, the critical chemical potential is
$\mu_{\rm crit}^{\rm 1L}({\cal G})=0.279 \sqrt{\cal G}$. Secondly, the
correction to the action produces an additional shift. We illustrate
this for the MN model though similar, (but less clean) conclusions can
be drawn from the Maris-Roberts type models \cite{FUTURE}.  Figure 1
(dashed line) shows the curve of critical chemical potential for the
one loop vertex MN model evaluated with the full action, ${\cal
  A}_{\rm 1L}$. Also illustrated in the bottom panel of this figure is
the effect of the correction term, ${\cal A}_{\Gamma}$, on the
critical chemical potential. Data are shown for the difference between
the critical chemical potentials obtained from the one loop vertex
solutions using the CJT and full, one loop actions; specifically for
$\delta\mu_{\rm crit}= [\mu^{\rm CJT}_{\rm crit}-\mu^{\rm 1L}_{\rm
  crit}]/\mu^{\rm 1L}_{\rm crit}$. It can be seen that the effect of
the correction term is insignificant ($|\delta\mu_{\rm crit}|<2$\% for
the range of values for $\cal G$ considered here) in comparison to the
modification of $\mu_{\rm crit}$ because of the use of one loop
solutions.


In summary, the studies performed here make it clear that the critical
chemical potential is an ``observable'' that is sensitive to the
non-perturbative structure of the quark self energy kernel. For
reasonable models with parameters fitted to the quark condensate, pion
mass and decay constant, we find that $\mu_{\rm crit}$ lies between
300 and 650~MeV.  Our studies utilising a one loop vertex support the
possibility that the CJT action, although not strictly appropriate,
provides a reliable extraction of $\mu_{\rm crit}$ for vertices other
than the bare vertex. Whether this remains true for (physically) more
acceptable vertices such as those of Ball and Chiu or Curtis and
Pennington\cite{BCCP} remains an open question.

\section*{Acknowledgement}
This work was supported by the Australian Research Council and
Adelaide University. We acknowledge helpful discussions with Reinhard
Alkofer, Peter Tandy and Craig Roberts.



\begin{table}
\begin{tabular}{||c|c|c|c|c|c|c||c||}
Model  & ${\cal G}$ [GeV$^2$] & $\omega$ [GeV]& $m_q [MeV]$ & 
$\left(-\langle \bar q q \rangle\right)^{1/3} [GeV]$ & $f_\pi$ [MeV] & $m_\pi$ [MeV]& $\mu_{\rm crit}$ [MeV] \\
\hline\hline
MN(bare)  & 1.0  & - & 17 & 0.216 & 93 & 138 & 340 \\
MN(bare)  & 1.20 & - & 17 & 0.236 & 102 & 145  & 370 \\
MN(loop)  & 0.93  & - & 15 & 0.211 & 93 & 137 & 270 \\
MN(loop)  & 1.15 & - & 15 & 0.236 & 105 & 144 & 300 \\ 
\hline
MR($\Delta=2$) & 0.82 & - & 9.8 & 0.236 & 90 & 138 & 630 \\
MR($\Delta=1$) & 0.61 & 0.3 & 6.6 & 0.236 & 77 & 138 & 610 \\
MR($\Delta=0$) & 0.45 & 0.3 & 5.7 & 0.236 & 75 & 139 & 530
\end{tabular}
\vspace*{0.5cm}
\caption{\label{param}
Parameters and observables are shown for a variety of models studied. 
The quark masses, $m_q$, listed for the Munczek-Nemirovsky (MN) models 
are the bare 
masses as this model is not renormalised. For the Maris-Roberts (MR) type 
models the value listed is the renormalised mass at 19 GeV$^2$. The 
different cases for each particular model are when either the condensate 
or pion decay constant is fitted.}
\end{table}


\begin{figure}
\begin{center}
  \epsfysize=11.6truecm \leavevmode \rotate[l]{\epsfbox{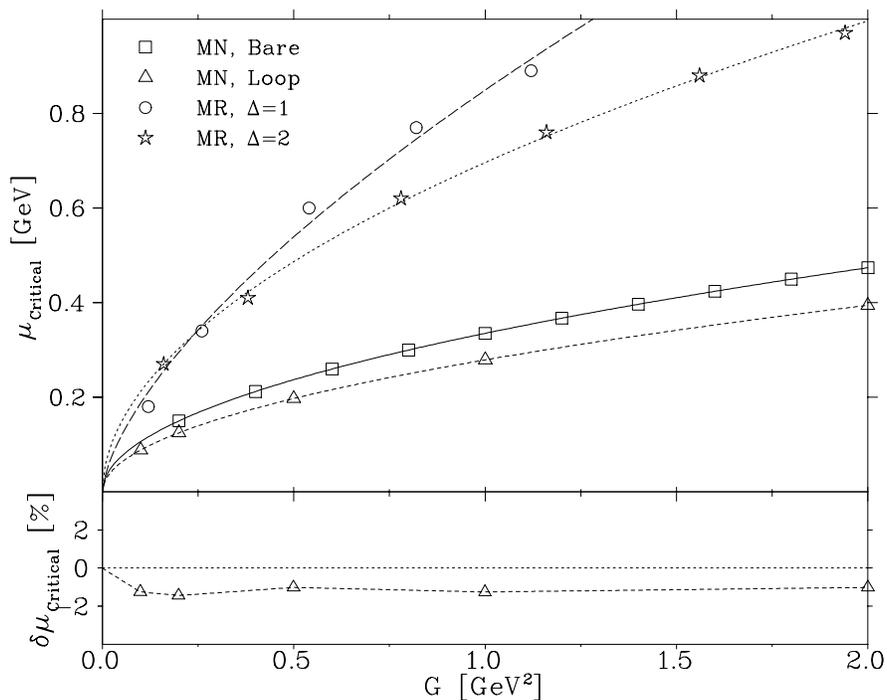}}
\caption{In the upper panel, the calculated critical chemical potential 
  is shown for the various self-energy kernels as a function of the
  coupling strength, ${\cal G}$ (G in axes labels). The cases are:
  (open squares) Munczek-Nemirovsky gluon propagator and bare vertex
  (CJT action); (open triangles) MN gluon and one-loop vertex
  (calculated using one-loop action); (open circles) Maris-Roberts
  (MR) type propagator, bare vertex with $\Delta=1$ and
  $\omega=0.3$~GeV; (open stars) MR propagator, bare vertex with
  $\Delta=2$.  Simple fits of the form $\mu_{\rm crit}({\cal
    G})=\alpha\,{\cal G}^\beta$ are shown for each data set. In the MN
  cases, the best fits occur with $\beta=1/2$ and with
  $\alpha=0.335(0.279)$ for the bare (loop) vertex.  With the MR
  propagator the best fits are: ($\Delta=2$ case) $\alpha=0.696$,
  $\beta=0.516$ and ($\Delta=1$ case) $\alpha=0.849$, $\beta=0.656$.
  In the lower panel, the effect on the critical chemical potential of
  using the corrected action for the one loop MN model solutions is
  shown.  This effect is small and the magnitude of $\delta\mu_{\rm
    crit}= [\mu^{\rm CJT}_{\rm crit}-\mu^{\rm 1L}_{\rm crit}]/\mu^{\rm
    1L}_{\rm crit}$ is less than $2$~\% over the range of $\cal G$
  studied.}
\label{phasestructure}
\end{center}
\end{figure}

\begin{figure}
\begin{center}  
  \epsfysize=11.6truecm \leavevmode
  \rotate[l]{\epsfbox{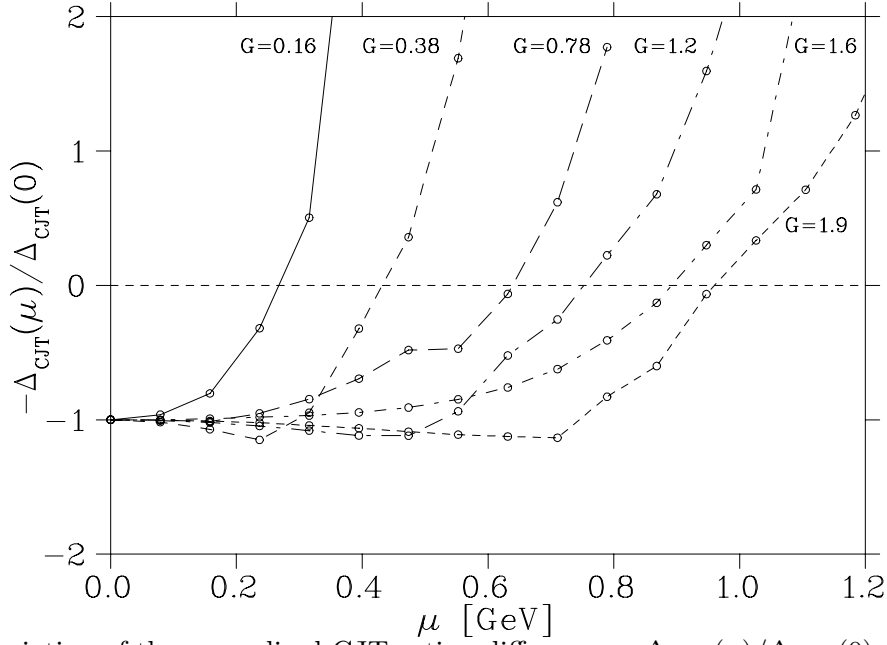}}
\caption{Variation of the normalised CJT action difference, 
  $-\Delta_{\rm CJT}(\mu)/\Delta_{\rm CJT}(0)$, with chemical
  potential is shown for a range of $\cal G$ (labelled in units of
  GeV$^2$). The critical chemical potential for a given $\cal G$ is
  reached at the zero intercept of that curve. }
\label{DeltavsG}
\end{center}
\end{figure}

\begin{figure}
\begin{center}  
  \epsfysize=11.6truecm \leavevmode \rotate[l]{\epsfbox{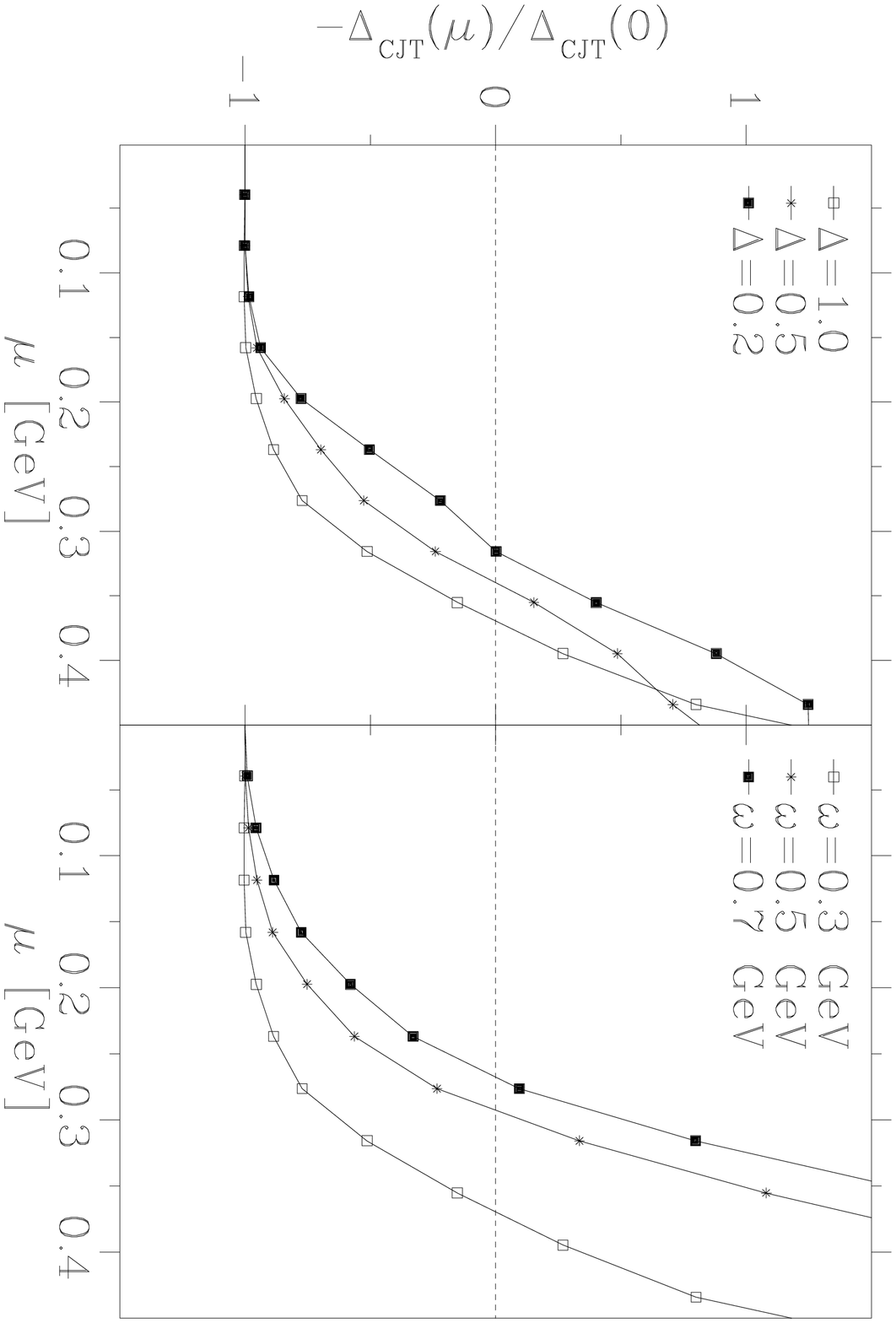}}
\caption{Variation of the normalised CJT action difference, 
  $-\Delta_{\rm CJT}(\mu)/\Delta_{\rm CJT}(0)$, with chemical
  potential is shown for a range of $\Delta$ (left panel, with fixed
  ${\cal G}=0.26$~GeV$^2$ and $\omega=0.3$~GeV) and $\omega$ (right
  panel, with $\Delta=1$ and fixed ${\cal G}=0.26$~GeV$^2$).}
\label{Deltavsomegadelta}
\end{center}
\end{figure}


\begin{thebibliography}{9}
  
  \bibitem{Karsch:2000vy} For a review, see F.~Karsch,
Nucl.\ Phys.\ Proc.\ Suppl.\ {\bf 83}, 14 (2000).
  
  \bibitem{Barbour:1998ej} I.~M.~Barbour, S.~E.~Morrison,
E.~G.~Klepfish, J.~B.~Kogut and M.~Lombardo,
Nucl.\ Phys.\ Proc.\ Suppl.\ {\bf 60A}, 220 (1998).

  \bibitem{DSErefs} For a recent review with emphasis on QCD thermodynamics, see 
C.~D.~Roberts and S.~M.~Schmidt,
Prog.\ Part.\ Nucl.\ Phys.\  {\bf 45S1}, 1 (2000)
[nucl-th/0005064];
Other useful reviews include 
R.~Alkofer and L.~von Smekal,
hep-ph/0007355.
 and
C.~D.~Roberts and A.~G.~Williams,
Prog.\ Part.\ Nucl.\ Phys.\ {\bf 33}, 477 (1994).
  
\bibitem{cscrefs}
K.~Rajagopal and F.~Wilczek,
hep-ph/0011333;
M.~Alford,
hep-ph/0102047.

    \bibitem{alford98} M.~Alford, K.~Rajagopal and F.~Wilczek,
  Phys.\ Lett.\ {\bf B422} (1998) 247 [hep-ph/9711395].

  \bibitem{Bloch:1999vk} J.~C.~Bloch, C.~D.~Roberts and S.~M.~Schmidt,
Phys.\ Rev.\ {\bf C60}, 065208 (1999).
  
  \bibitem{Bender:1998jf} A.~Bender, G.~I.~Poulis, C.~D.~Roberts,
S.~M.~Schmidt and A.~W.~Thomas,
Phys.\ Lett.\ {\bf B431}, 263 (1998).
    
\bibitem{Hawes:1998cw}
F.~T.~Hawes, P.~Maris and C.~D.~Roberts,
Phys.\ Lett.\ B {\bf 440}, 353 (1998)
  
  \bibitem{Munczek:1983dx} H.~J.~Munczek and A.~M.~Nemirovsky,
Phys.\ Rev.\ {\bf D28}, 181 (1983).
  
  \bibitem{Blaschke:1998bj} D.~Blaschke, C.~D.~Roberts and
S.~M.~Schmidt,
Phys.\ Lett.\ {\bf B425}, 232 (1998.
  
  \bibitem{Bonnet:2000kw} F.~D.~Bonnet, P.~O.~Bowman, D.~B.~Leinweber
and A.~G.~Williams,
Phys.\ Rev.\ D {\bf 62}, 051501 (2000)

  \bibitem{GluonDSE} P.~Watson and R.~Alkofer,
hep-ph/0102332;
R.~Alkofer and L.~von Smekal,
hep-ph/0007355;
D.~Atkinson and J.~C.~Bloch,
Mod.\ Phys.\ Lett.\ A {\bf 13}, 1055 (1998).
L.~von Smekal, R.~Alkofer and A.~Hauck,
Phys.\ Rev.\ Lett.\ {\bf 79}, 3591 (1997);

  \bibitem{Frank:1996uk} M.~R.~Frank and C.~D.~Roberts,
Phys.\ Rev.\ {\bf C53}, 390 (1996).
  
  \bibitem{Maris:1997tm} P.~Maris and C.~D.~Roberts,
Phys.\ Rev.\ {\bf C56}, 3369 (1997).
  
  \bibitem{Maris:1999nt} P.~Maris and P.~C.~Tandy,
Phys.\ Rev.\ {\bf C60}, 055214 (1999).
  
  \bibitem{Cornwall:1974vz} J.~M.~Cornwall, R.~Jackiw and
E.~Tomboulis,
Phys.\ Rev.\ {\bf D10}, 2428 (1974).
      
  \bibitem{BCCP} J.~S.~Ball and T.~Chiu,
Phys.\ Rev.\ {\bf D22}, 2542 (1980);
Phys.\ Rev.\ {\bf D22}, 2550 (1980); 
D.~C.~Curtis and M.~R.~Pennington,
Phys.\ Rev.\ {\bf D42}, 4165 (1990).
  
  \bibitem{FUTURE} A.~Bender, W.~Detmold, A.~W.~Thomas, {\it in
  preparation}.

  \bibitem{Bender:1996bb} A.~Bender, C.~D.~Roberts and L.~von Smekal,
Phys.\ Lett.\ {\bf B380}, 7 (1996).
  
  \bibitem{Hellstern:1997nv} G.~Hellstern, R.~Alkofer and
H.~Reinhardt,
Nucl.\ Phys.\ {\bf A625}, 697 (1997).
    
\end{thebibliography}
\end{document}